\documentclass[twocolumn,showpacs,preprintnumbers,amsmath,amssymb,prb]{revtex4}

\usepackage{graphicx}
\usepackage{dcolumn}
\usepackage{bm}
\usepackage{color}

\begin{document}

\preprint{}

\title{Acoustic study of dynamical molecular-spin state without undergoing magnetic phase transition in spin-frustrated ZnFe$_2$O$_4$}

\author{Tadataka Watanabe$^1$}
\thanks{tadataka@phys.cst.nihon-u.ac.jp}
\author{Shota Takita$^1$}
\author{Keisuke Tomiyasu$^2$}
\author{Kazuya Kamazawa$^3$}
\affiliation{$^1$Department of Physics, College of Science and Technology (CST), Nihon University, Chiyoda, Tokyo 101-8308, Japan}

\affiliation{$^2$Department of Physics, Tohoku University, Sendai, Miyagi 980-8577, Japan}
\affiliation{$^3$Comprehensive Research Organization for Science and Society (CROSS), Tokai, Ibaraki 319-1106, Japan}
\date{\today}

\begin{abstract}
Ultrasound velocity measurements were performed on a single crystal of spin-frustrated ferrite spinel ZnFe$_2$O$_4$ from 300 K down to 2 K. In this cubic crystal, all the symmetrically-independent elastic moduli exhibit softening with a characteristic minimum with decreasing temperature below $\sim$100 K. This elastic anomaly suggests a coupling between dynamical lattice deformations and molecular-spin excitations. In contrast, the elastic anomalies, normally driven by the magnetostructural phase transition and its precursor, are absent in ZnFe$_2$O$_4$, suggesting that the spin-lattice coupling cannot play a role in relieving frustration within this compound. The present study infers that, for ZnFe$_2$O$_4$, the dynamical molecular-spin state evolves at low temperatures without undergoing precursor spin-lattice fluctuations and spin-lattice ordering. It is expected that ZnFe$_2$O$_4$ provides the unique dynamical spin-lattice liquid-like system, where not only the spin molecules but also the cubic lattice fluctuate spatially and temporally.
\end{abstract}


\maketitle

\section{Introduction}

Cubic spinels $AB_2$O$_4$ with magnetic $B$ ions have attracted considerable interest in light of the geometrical frustration which is inherent in the $B$-site sublattice of corner-sharing tetrahedra (pyrochlore lattice).\cite{Lee3} One of the most extensively studied spinel systems is chromite spinels $A$Cr$_2$O$_4$ with $A$ = Mg and Zn, for which the magnetic properties are fully dominated by the Jahn-Teller (JT)-inactive Cr$^{3+}$ with spin $S$ = 3/2 (Fig.~1(a)) residing on the pyrochlore network.\cite{Ueda} $A$Cr$_2$O$_4$ with Weiss temperature $\Theta_W\simeq-$390~K undergoes an antiferromagnetic (AF) ordering at $T_N\simeq$ 13~K along with a cubic-to-tetragonal structural distortion.\cite{Lee,Ortega-San-Martin,Chung} Ferrite spinels $A$Fe$_2$O$_4$ with $A$ = Zn and Cd are another JT-inactive spinel system with Fe$^{3+}$ showing a high spin of $S$ = 5/2 (Fig.~1(b)).\cite{Schiessl} For $A$Fe$_2$O$_4$ with $\Theta_W\simeq120$~K ($A$ = Zn) and $\simeq-$50~K ($A$ = Cd), neutron scattering experiments in the high-purity single crystals observed neither long-range magnetic ordering nor a structural transition down to low temperature (1.5~K) although an AF-transition-like anomaly occurs in the magnetic susceptibility at $T^*\simeq13$~K (Fig.~1(c)).\cite{Kamazawa,Kamazawa2} Additionally, it is noted that the magnetic susceptibility of ZnFe$_2$O$_4$ exhibits a deviation from the Curie-Weiss law below $\sim$100 K (Fig.~1(d)),\cite{Kamazawa} which implies the enhancement of the AF interactions at low temperatures.\cite{Yamada} Thus, the frustrated magnetism of $A$Fe$_2$O$_4$ should be different in nature from that of $A$Cr$_2$O$_4$.

For $A$Cr$_2$O$_4$, the phase transition to spin-lattice ordering is explained by the spin-JT mechanism via spin-lattice coupling, where local distortions of the tetrahedra release the frustration in the nearest-neighbor AF interactions.\cite{Yamashita,Tchernyshyov,Ji} In the frustrated paramagnetic (PM) phase of $A$Cr$_2$O$_4$, inelastic neutron scattering (INS) experiments provided evidence of quasielastic magnetic scattering, indicating the presence of strong spin fluctuations because of spin frustration.\cite{Lee,Lee2,Suzuki,Tomiyasu} This quasielastic mode involved the fluctuations of AF hexagonal spin molecules (AF hexamers) in the pyrochlore lattice (Fig.~1(e)).\cite{Lee2,Tomiyasu} Further, ultrasound velocity measurements of $A$Cr$_2$O$_4$ suggested the coexistence/crossover of the precursor spin-lattice fluctuations towards a phase transition (spin-JT fluctuations) and the gapped molecular-spin excitations also coupled with the lattice,\cite{Kino,Watanabe} which is compatible with the recent observation of finite-energy molecular-spin excitations in time-of-flight INS experiments in the PM phase of this compound.\cite{Tomiyasu2}

For $A$Fe$_2$O$_4$, whereas spin-lattice ordering is absent down to low temperature, the INS experiments observed magnetic diffuse scattering and its very soft dispersion relation in the energy range below $\sim$2 meV, arising possibly from the dynamical molecular-spin state.\cite{Kamazawa,Kamazawa2} Thus, in the absence of the spin-JT effect, the frustrated magnetism in $A$Fe$_2$O$_4$ is expected to be mainly governed by the dynamical molecular-spin state. For ZnFe$_2$O$_4$, the observed diffuse scattering was attributed to the fluctuations of AF twelve-membered spin molecules (AF dodecamers illustrated in Fig.~1(f)).\cite{Tomiyasu3} The formation of the different types of spin molecules in between ZnFe$_2$O$_4$ (the AF dodecamers) and $A$Cr$_2$O$_4$ (the AF hexamers) is considered to arise from the difference in the dominant exchange paths, specifically, the third-neighbor AF interactions $J_3$ with additional nearest-neighbor ferromagnetic (FM) $J_1$ for ZnFe$_2$O$_4$,\cite{Kamazawa,Tomiyasu3} but AF $J_1$ for $A$Cr$_2$O$_4$,\cite{Lee2,Tomiyasu}, respectively. For CdFe$_2$O$_4$, the INS experiments produced scattering patterns, which resembles that of $A$Cr$_2$O$_4$, indicative of the dominant AF $J_1$.\cite{Kamazawa2}

Interestingly, the diffuse-neutron-scattering patterns of ZnFe$_2$O$_4$ vary with temperature,\cite{Kamazawa} while those of $A$Cr$_2$O$_4$ and CdFe$_2$O$_4$ are independent of temperature.\cite{Kamazawa2,Lee2,Tomiyasu} This temperature dependence in ZnFe$_2$O$_4$ was explained by the competition between the third-neighbor AF $J_3$ and the temperature-dependent nearest-neighbor FM $J_1$, where the nearest-neighbor FM $J_1$ are weakened with decreasing temperature due to the temperature-dependent bond angle of the nearest-neighbor Fe$^{3+}$-O-Fe$^{3+}$.\cite{Kamazawa,Yamada} The neutron scattering experiments in ZnFe$_2$O$_4$ suggested that the deviation from the Curie-Weiss law below $\sim$100 K in the magnetic susceptibility of this compound (Fig.~1(d)) arises from the AF component, whereas the FM component leads to the Curie-Weiss behavior with $\Theta_W\simeq$ 120 K at high temperatures.\cite{Kamazawa} And it is considered that, for ZnFe$_2$O$_4$, the formation of the spin molecules (the AF dodecamers) is realized at low temperatures, where the AF component generates the frustration.\cite{Tomiyasu3}

\begin{figure}[t]
\begin{center}
\includegraphics[scale=0.4]{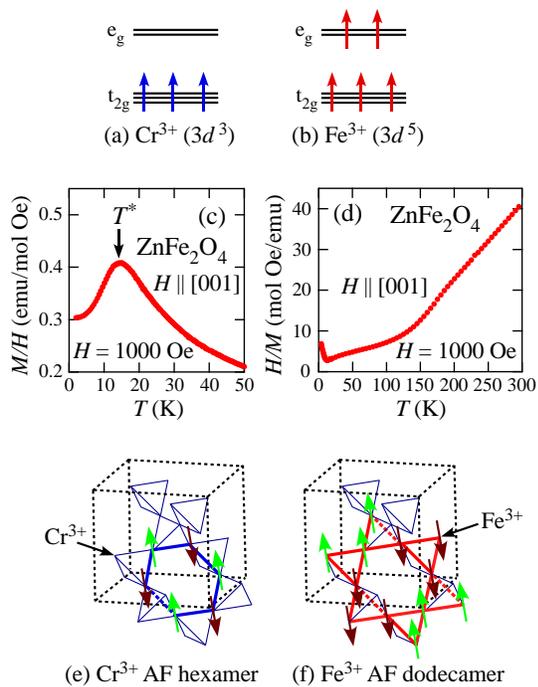}
\caption{\label{fig:fig1} (Color online) (a) and (b) show the spin states of Cr$^{3+}$ (3$d^3$) and high-spin Fe$^{3+}$ (3$d^5$) in the octahedral crystal field, respectively. (c) and (d) depict respectively the magnetic susceptibility and the inversed magnetic susceptibility of single-crystalline ZnFe$_2$O$_4$ as functions of temperature. (e) and (f) illustrate respectively a Cr$^{3+}$ AF hexamer and a Fe$^{3+}$ AF dodecamer in the pyrochlore lattice. The dotted cube in (e) and (f) depicts a unit cell of the cubic spinel structure.}
\end{center}
\end{figure}

Notably, molecular-spin excitations were observed in the INS experiments in the frustrated magnets of not only $A$Cr$_2$O$_4$ and $A$Fe$_2$O$_4$ but also HgCr$_2$O$_4$,\cite{Tomiyasu4} GeCo$_2$O$_4$,\cite{Tomiyasu5} LiV$_2$O$_4$,\cite{Tomiyasu6} and Tb$_2$Ti$_2$O$_7$,\cite{Yasui} where the number of magnetic ions, shape, and symmetry of spin molecules are considered to vary from compound to compound depending on the dominant exchange path. These observations imply that the dynamical molecular-spin state can universally emerge in the frustrated magnets. Thus a comparative study among the spinel magnets using a variety of experimental probes must provide a root for understanding the nature of dynamical molecular-spin state.

In this paper, we present an analysis of ultrasound velocity measurements for the zinc ferrite spinel ZnFe$_2$O$_4$. The sound velocity or the elastic modulus is a useful probe enabling symmetry-resolved thermodynamic information to be extracted from frustrated magnets.\cite{Kino,Watanabe,Kino2,Zherlitsyn,Wolf,Watanabe2,Bhattacharjee,Watanabe3,Felea,Nii,Watanabe4} As mentioned earlier, the observed elastic anomalies in the chromite spinels $A$Cr$_2$O$_4$ inferred the coexistence of spin-JT fluctuations and molecular-spin excitations in the PM phase. \cite{Kino,Watanabe} The present study finds elastic anomalies in ZnFe$_2$O$_4$ that suggest the evolution of a dynamical molecular-spin state at low temperatures without undergoing precursor spin-lattice fluctuations and spin-lattice ordering, which is a behavior uniquely different from another spin-frustrated molecular-spin system $A$Cr$_2$O$_4$. Moreover, our study also suggests that, for ZnFe$_2$O$_4$, the molecular-spin excitations consist of multiple gapped modes and sensitively couple to the trigonal lattice deformations, which is a behavior similar to $A$Cr$_2$O$_4$ although the details are different in between ZnFe$_2$O$_4$ and $A$Cr$_2$O$_4$.

\section{Experimental}

The ultrasound velocity measurements were performed on a high-purity single crystal of ZnFe$_2$O$_4$ grown by the flux method.\cite{Kamazawa} Figure 1(c) and (d) plots respectively the temperature dependence of the magnetic susceptibility and the inversed magnetic susceptibility of the ZnFe$_2$O$_4$ single crystal used in the present study, where the AF-transition-like anomaly at $T^*\sim$13~K and the deviation from the Curie-Weiss law below $\sim$100 K occur.\cite{Kamazawa} The ultrasound velocities were measured using the phase-comparison technique with longitudinal and transverse sound waves at a frequency of 30 MHz. The ultrasonic waves were generated and detected by LiNbO$_3$ transducers glued to the parallel mirror surfaces of the crystal. The measurements were performed at temperatures from 300~K to 2~K for all the symmetrically-independent elastic moduli in the cubic crystal, specifically, compression modulus $C_{11}$, tetragonal shear modulus $\frac{C_{11}-C_{12}}{2}\equiv C_t$, and trigonal shear modulus $C_{44}$. The respective measurements of $C_{11}$, $C_t$, and $C_{44}$ were performed using longitudinal sound waves with propagation {\bf k}$\parallel$[001] and polarization {\bf u}$\parallel$[001], transverse sound waves with {\bf k}$\parallel$[110] and {\bf u}$\parallel$[1$\bar{1}$0], and transverse sound waves with {\bf k}$\parallel$[110] and {\bf u}$\parallel$[001]. The sound velocities of ZnFe$_2$O$_4$ measured at room temperature (300~K) are 6480 m/s for $C_{11}$, 2930 m/s for $C_t$, and 3740 m/s for $C_{44}$.

\section{Results And Discussion}

Figure 2(a)-(c) presents the temperature dependence of the elastic moduli, $C_{11}(T)$, $C_t(T)$, and $C_{44}(T)$ in ZnFe$_2$O$_4$. On cooling from room temperature (300~K) to $\sim$100~K, all the elastic moduli exhibit ordinary hardening consistent with the background $C_{\Gamma}^0(T)$ taken from an empirical evaluation of the experimental $C_{\Gamma}(T)$ in 100 K $<T<$ 300 K (dotted curves in Fig.~2(a)-(c)) \cite{Varshni}. Here, the background values at $T$ = 0 K, $C_{11}^0(0)\simeq$ 233.8 GPa, $C_t^0(0)\simeq$ 46.7 GPa, and $C_{44}^0(0)\simeq$ 75.9 GPa give the respective sound velocities of $v_{11}^0\simeq$ 6620 m/s, $v_t^0\simeq$ 2960 m/s, and $v_{44}^0\simeq$ 3770 m/s, which are $\sim$ 2 $\%$, $\sim$ 1 $\%$, and $\sim$ 0.8 $\%$ larger than the measured sound velocities at 300 K, respectively. The values of $v_{11}^0$, $v_t^0$, and $v_{44}^0$ give the averaged sound velocity $\bar{v}\sim$ 3700 m/s,\cite{Luthi} which is compatible with the Debye temperature $\Theta_D\sim$ 250 K for ZnGa$_2$O$_4$, a nonmagnetic reference compound for ZnFe$_2$O$_4$, giving the averaged sound velocity $\bar{v}\sim$ 3500 m/s.\cite{Martinho}

In Fig. 2(a)-(c), below $\sim$100~K, all the elastic moduli of ZnFe$_2$O$_4$ exhibit an anomalous temperature variation, specifically, the deviation from the ordinal hardening indicated as the dotted curves in Fig. 2(a)-(c). $C_{11}(T)$ exhibits softening with decreasing temperature below $\sim$50~K but turns to hardening below $\sim$6~K. Similarly, $C_t(T)$ and $C_{44}(T)$ exhibit softening with decreasing temperature below $\sim$80~K, which turn to hardening below $\sim$6~K and $\sim$4~K, respectively. Taking into account the absence of orbital degeneracy in the $B$-site high-spin Fe$^{3+}$ in ZnFe$_2$O$_4$ (Fig. 1(b)), the anomaly in $C_{\Gamma}(T)$ should have a magnetic origin where the spin degrees of freedom play a significant role.\cite{Schiessl} Note that the elastic anomalies emerge at temperatures where the magnetic susceptibility exhibits the deviation from the Curie-Weiss law (Fig.~1(d)). This correspondence implies that the elastic anomalies are driven by the generation of frustration below $\sim$100 K, which is compatible with the temperature-dependent $J_1/J_3$ suggested from the neutron scattering experiments.\cite{Kamazawa}

\begin{figure}[t]
\begin{center}
\includegraphics[scale=0.4]{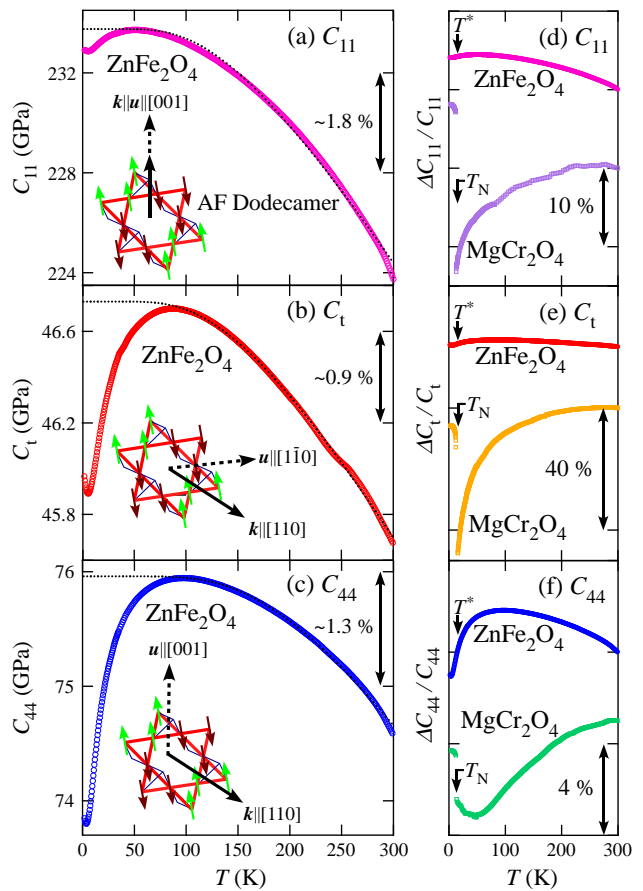}
\caption{\label{fig:fig2_revised} (Color online) (a)-(c) Elastic moduli of ZnFe$_2$O$_4$ as functions of temperature. (a) $C_{11}(T)$, (b) $C_t(T)$, and (c) $C_{44}(T)$. The dotted curves in (a)-(c) indicate the background $C^0_{\Gamma}(T)$ in each modulus taken from an empirical evaluation of the experimental $C_{\Gamma}(T)$ in 100 K $<T<$ 300 K.\cite{Varshni} The insets to (a)-(c) illustrate single AF dodecamers in the Fe$^{3+}$ pyrochlore lattice with the propagation {\bf k} and polarization {\bf u} of sound waves in the respective elastic modes. (d)-(f) The dependence on temperature of the elastic moduli of ZnFe$_2$O$_4$ [from Fig.~2(a)-(c)] and MgCr$_2$O$_4$.\cite{Watanabe} (d) $C_{11}(T)$, (e) $C_t(T)$, and (f) $C_{44}(T)$. The curves are vertically shifted for clarity. $T^*$ and $T_N$ in (d)-(f) indicate the temperature at which the AF-transition-like anomaly occurs in the magnetic susceptibility for ZnFe$_2$O$_4$ seen in Fig.~1(c), and the AF ordering temperature for MgCr$_2$O$_4$, respectively.}
\end{center} 
\end{figure}

The elastic anomalies in the JT-inactive magnets like ZnFe$_2$O$_4$ are attributed to magnetoelastic coupling acting on the exchange interactions. In this mechanism, the exchange striction arises from a modulation of the exchange interactions by ultrasonic waves.\cite{Luthi} Both the longitudinal and transverse sound waves couple to the spin system via the exchange striction mechanism, which depends on the directions of both polarization {\bf u} and propagation {\bf k} of sound waves relative to the exchange path.

Similar to ZnFe$_2$O$_4$, the softening-with-minimum elastic anomaly in $C_{\Gamma}(T)$ is also observed in other frustrated magnets of $A$Cr$_2$O$_4$,\cite{Kino,Watanabe} SrCu$_2$(BO$_3$)$_2$,\cite{Zherlitsyn,Wolf} GeCo$_2$O$_4$,\cite{Watanabe3} and MgV$_2$O$_4$,\cite{Watanabe4} the origin of which is considered to be the coupling of the lattice to the gapped magnetic excitations via the exchange striction mechanism. Recalling that molecular-spin excitations, i.e., the excitations of AF dodecamers, were observed in the INS experiments for ZnFe$_2$O$_4$,\cite{Kamazawa,Tomiyasu3} the softening-with-minimum exhibited in $C_{\Gamma}(T)$ for ZnFe$_2$O$_4$ arises from the coupling of the lattice to the molecular-spin excitations, which is similar to the softening-with-minimum observed in $C_{\Gamma}(T)$ of $A$Cr$_2$O$_4$.\cite{Kino,Watanabe} The insets to Fig.~2(a)-(c) illustrate single AF dodecamers in the Fe$^{3+}$ pyrochlore lattice with the propagation {\bf k} and polarization {\bf u} of sound waves in the respective elastic modes. From the symmetry point of view, the AF-dodecamer excitations should couple more sensitively to the trigonal lattice deformations generated by sound waves with {\bf k}$\parallel$[110] and {\bf u}$\parallel$[001] (the inset to Fig.~2(c)). For ZnFe$_2$O$_4$, as shown in Fig. 2(a)-(c), the magnitude of the softening in $C_{\Gamma}(T)$ is indeed largest in the trigonal shear modulus $C_{44}(T)$, specifically, $\Delta C_{11}/C_{11}\sim 0.4$ $\%$, $\Delta C_t/C_t\sim 1.8$ $\%$, and $\Delta C_{44}/C_{44}\sim 2.8$ $\%$, which is compatible with the trigonal symmetry of the AF dodecamer.

Figure 2(d)-(f) compares the relative shifts of $C_{11}(T)$, $C_{t}(T)$, and $C_{44}(T)$, respectively, in between ZnFe$_2$O$_4$ [from Fig.~2(a)-(c)] and MgCr$_2$O$_4$.\cite{Watanabe} Note here that $C_{\Gamma}(T)$ for MgCr$_2$O$_4$ in the PM phase ($T>T_N$) exhibits not only softening-with-minimum in $C_{44}(T)$ from molecular-spin excitations but also a huge Curie-type $-1/T$ softening in $C_{11}(T)$ and $C_t(T)$, being a precursor to the spin-JT transition.\cite{Watanabe} This coexistence of two types of elastic anomalies in MgCr$_2$O$_4$ infers the coexistence of molecular-spin excitations and spin-JT fluctuations. For ZnFe$_2$O$_4$, in contrast, $C_{\Gamma}(T)$ exhibits solely softening-with-minimum, as is clearly seen in the expanded view of $C_{\Gamma}(T)$ (Fig.~3(a)-(c) [open circles, from Fig.~2(a)-(c)]), which infers the presence of molecular-spin excitations but the absence of spin-JT fluctuations.

We also note here that, whereas $C_{\Gamma}(T)$ for MgCr$_2$O$_4$ exhibits a discontinuity at $T_N$ (Fig.~2(d)-(f)), $C_{\Gamma}(T)$ for ZnFe$_2$O$_4$ exhibits no discontinuity at $T^*$ (Figs.~2(d)-(f) and 3(a)-(c)). Thus, the experimental $C_{\Gamma}(T)$ for ZnFe$_2$O$_4$ indicates that the AF-transition-like anomaly at $T^*$ in the magnetic susceptibility seen in Fig.~1(c) is not a phase transition. This inference is compatible with the absence of long-range magnetic ordering at least down to 1.5 K as revealed by the neutron scattering experiments.\cite{Kamazawa} Regarding $C_{\Gamma}(T)$ for ZnFe$_2$O$_4$, the continuity in elasticity at $T^*$ (the absence of a phase transition) is compatible with the absence of Curie-type softening (the absence of a precursor for the magnetostructural transition), indicating that the spin-lattice coupling cannot produce the spin-JT transition because the strength of the exchange interactions is not large enough to overcome the cost in elastic energy involved in the static long-range lattice deformation. As a result, it is considered that, for ZnFe$_2$O$_4$, solely the dynamical molecular-spin state emerges without undergoing spin-lattice fluctuations and spin-lattice ordering, which is different from the coexistence of the dynamical molecular-spin state and the spin JT effect in $A$Cr$_2$O$_4$. However, the precise nature of the magnetic state of ZnFe$_2$O$_4$ below $T^*$ remains to be discovered. Freezing of the spin molecules is a possibility that might occur at $T^*$ in ZnFe$_2$O$_4$. Furthermore, although the spin-lattice coupling in ZnFe$_2$O$_4$ is much weaker than that in $A$Cr$_2$O$_4$ as indicated in Fig. 2(d)-(f), there remains a possibility of spin-JT transition at low $T<$ 1.5 K for ZnFe$_2$O$_4$.

As is clear from a comparison between $C_{\Gamma}(T)$ for ZnFe$_2$O$_4$ (Fig.~2(a)-(c)) and $C_{44}(T)$ for MgCr$_2$O$_4$ (Fig.~2(f)), the softening in $C_{\Gamma}(T)$ begins to occur below $\sim$50~K or $\sim$80~K in ZnFe$_2$O$_4$ but above 300~K in MgCr$_2$O$_4$. The softening occurring at lower temperatures for ZnFe$_2$O$_4$ indicates the evolution of the dynamical molecular-spin state at lower temperatures, which is driven by the generation of frustration below $\sim$100~K because of the temperature-dependent $J_1/J_3$. Furthermore, the magnitude of the softening in $C_{\Gamma}(T)$ for ZnFe$_2$O$_4$ is smaller than that in $C_{44}(T)$ for MgCr$_2$O$_4$. As discussed later in conjunction with Eq.~(1) and Table I, the reason for the smaller magnitude of the softening for ZnFe$_2$O$_4$ relative to MgCr$_2$O$_4$ is because the coupling is weaker between the dynamical lattice deformations and molecular-spin excitations. Additionally, as is also clear from a comparison between $C_{\Gamma}(T)$ of ZnFe$_2$O$_4$ (Fig.~3(a)-(c)) and $C_{44}(T)$ of MgCr$_2$O$_4$ (Fig.~2(f)), the former exhibits its minimum at $\sim$5~K, which is lower than the minimum point of $\sim$50~K in the latter. As also discussed later in conjunction with Eq.~(1) and Table I, the lower temperature at which the minimum point occurs for ZnFe$_2$O$_4$ relative to MgCr$_2$O$_4$ is due to the smaller gap associated with its molecular-spin excitations.

Softening-with-minimum in $C_{\Gamma}(T)$ driven by the molecular-spin excitations is generally explained as the presence of a finite gap for the excitations, which is sensitive to strain.\cite{Watanabe} In the mean-field approximation, the elastic modulus $C_{\Gamma}(T)$ of the molecular-spin system is written as \cite{Watanabe}:
\begin{equation}
C_{\Gamma}(T)=C^0_{\Gamma}(T)-G_{1,\Gamma}^2N\frac{\chi_{\Gamma}(T)}{\left\{1-K_{\Gamma}\chi_{\Gamma}(T)\right\}},
\label{eq:SM}
\end{equation}
where $C^0_{\Gamma}(T)$ is the background elastic constant, $N$ the density of spin molecules, $G_{1,\Gamma}=|\partial \Delta_1/\partial \epsilon_{\Gamma}|$ the coupling constant for a single spin molecule measuring the strain ($\epsilon_{\Gamma}$) dependence of the excitation gap $\Delta_1$, $K_{\Gamma}$ the inter-spin-molecule interaction, and $\chi_{\Gamma}(T)$ the strain susceptibility of a single spin molecule. From Eq.~(1), the minimum in $C_{\Gamma}(T)$ appears when this elastic mode strongly couples to the excited state at $\Delta_1$; on cooling, $C_{\Gamma}(T)$ exhibits softening roughly down to $T\sim\Delta_1$, but recovery of the elasticity (hardening) roughly below $T\sim\Delta_1$.

As explained above, the softening-with-minimum anomaly in $C_{\Gamma}(T)$ for ZnFe$_2$O$_4$ should arise from a gap in the molecular-spin excitations that is sensitive to the strain. This interpretation helps to understand the INS results for ZnFe$_2$O$_4$.\cite{Kamazawa} The broad magnetic scattering spectrum should observe gapped molecular-spin excitations that are considerably smeared. The smeared INS spectra at 1.5 K, only one-third of the minimum position of $\sim$5 K in $C_{\Gamma}(T)$, are probably due to strong spin frustration as well as some kind of quantum effect in the molecular-spin system. Recall that, for other frustrated magnets of $A$Cr$_2$O$_4$ and SrCu$_2$(BO$_3$)$_2$, observations have been reported of the softening-with-minimum in $C_{\Gamma}(T)$ and the $T$-dependent observation of the gapped magnetic excitations in the INS spectra. For $A$Cr$_2$O$_4$, whereas $C_{\Gamma}(T)$ exhibited the minimum at $\sim$50 K [Fig.~2(f)],\cite{Watanabe} the INS experiments observed broad quasielastic magnetic scattering spectrum at temperatures down to $T_N\simeq$ 13 K but observed distinct $\sim$4-meV gapped excitations below $T_N$.\cite{Lee,Lee2,Tomiyasu} For the dimer-spin system SrCu$_2$(BO$_3$)$_2$, softening-with-minimum in $C_{\Gamma}(T)$ was observed,\cite{Zherlitsyn,Wolf} whereas the INS experiments demonstrated $\sim$3-meV gapped excitations at temperatures only below $\sim$10 K.\cite{Kageyama} For ZnFe$_2$O$_4$, it is expected that the INS experiments at low $T<$ 1.5 K show clear gapped molecular-spin excitations.

\begin{figure}[t]
\begin{center}
\includegraphics[scale=0.4]{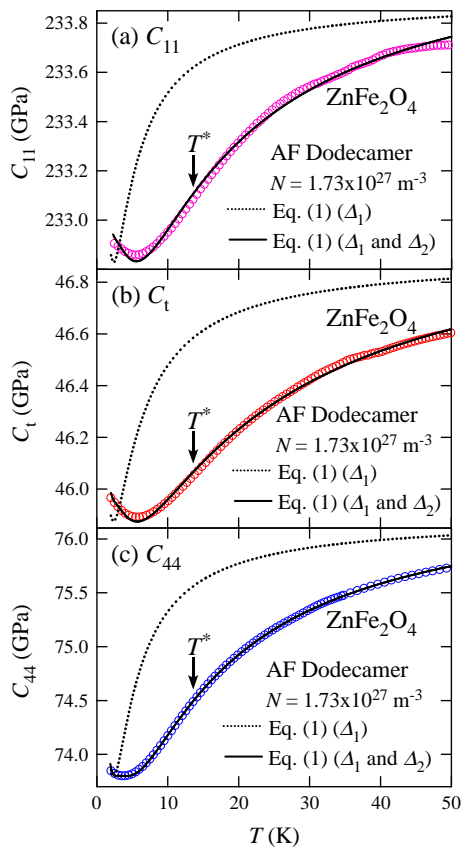}
\caption{\label{fig:fig3} (Color online) Expanded view of $C_{\Gamma}(T)$ below 50~K in ZnFe$_2$O$_4$ [open circles, from Fig.~2(a)-(c)]. (a) $C_{11}(T)$, (b) $C_t(T)$, and (c) $C_{44}(T)$. $T^*$ in (a)-(c) indicates the temperature at which the AF-transition-like anomaly occurs in the magnetic susceptibility for ZnFe$_2$O$_4$ seen in Fig.~1(c). The solid curves (dotted curves) in (a)-(c) are fits of the experimental $C_{\Gamma}(T)$ to Eq.~(1) with two singlet-triplet gaps $\Delta_1$ and $\Delta_2$ (single singlet-triplet gap $\Delta_1$) for the AF-dodecamer excitations.\cite{Tomiyasu3}}
\end{center}
\end{figure}

We now give a quantitative analysis of the experimental $C_{\Gamma}(T)$ in ZnFe$_2$O$_4$ using Eq.~(1) assuming excitations of the AF dodecamers in the Fe$^{3+}$ pyrochlore lattice [Fig.~1(f)].\cite{Tomiyasu3} Here, the value of the density of AF dodecamers, $N$ in Eq.~(1), is assumed to be $N$ = 1.73$\times$10$^{27}$ m$^{-3}$, which is one-twelfth of the density of Fe$^{3+}$ ions in ZnFe$_2$O$_4$.\cite{Tomiyasu3} We fit Eq.~(1) to the experimental data $C_{\Gamma}(T)$ below 50~K. Although the background $C^0_{\Gamma}(T)$ in Eq.~(1) generally exhibits hardening with decreasing temperature,\cite{Varshni} we here assume that $C^0_{\Gamma}(T)$ is constant because, for ZnFe$_2$O$_4$, the hardening of the background below $\sim$50~K is negligibly small compared with the softening-with-minimum in $C_{\Gamma}(T)$, as indicated in Fig.~2(a)-(c).

\begin{table}[t]
\caption{\label{tab:table1} Values of the fitting parameters in Eq.~(1) with two singlet-triplet gaps $\Delta_1$ and $\Delta_2$ for the experimental $C_{\Gamma}(T)$ of ZnFe$_2$O$_4$ [from Fig.~3(a)-(c)] (upper column) and $A$Cr$_2$O$_4$ ($A$ = Mg and Zn) (lower column).\cite{Watanabe} For ZnFe$_2$O$_4$ (upper column), values of the fitting parameters in Eq.~(1) with singlet-triplet $\Delta_1$ and singlet-nonet $\Delta_2$ are also shown in parentheses.}
\begin{ruledtabular}
\begin{tabular}{ccccccc}
& & $\Delta_1$ (K) & $G_{1,\Gamma}$ (K) & $\Delta_2$ (K) & $G_{2,\Gamma}$ (K) & $K_{\Gamma}$ (K)\\
\hline
ZnFe$_2$O$_4$ & $C_{11}$ & & 890 & & 2630 & -6 \\
$[$Dodec.$]$ & & & (1190) & & (3180) & (-18) \\
 & $C_t$ & 5 & 980 & 15 & 2910 & -10 \\
 & & (5) & (1230) & (15) & (3470) & (-23) \\
 & $C_{44}$ & & 1350 & & 3570 & -6 \\
 & & & (2038) & & (4370) & (-27) \\
\hline
MgCr$_2$O$_4$ & $C_{44}$ & 39 & 3600 & 136 & 10200 & -19 \\
ZnCr$_2$O$_4$ & $C_{44}$ & 34 & 3160 & 111 & 9290 & -19 \\
$[$Hex.$]$ & & & & & & \\
\end{tabular}
\end{ruledtabular}
\end{table}

For ZnFe$_2$O$_4$, taking into account the vanishing total spin $S_{tot}=0$ in the ground state of the AF dodecamers,\cite{Tomiyasu3} $\Delta_1$ in Eq.~(1) is assumed to be singlet-multiplet excitations of the single AF dodecamers. We note that, to reproduce the softening in the experimental $C_{\Gamma}(T)$ for ZnFe$_2$O$_4$ using Eq.~(1), we must assume the coupling of the lattice to not only the lowest excitations $\Delta_1$ but also the higher excitations $\Delta_i$ ($i$ = 2, 3, 4, $\ldots$). Dotted curves in Fig. 3(a)-(c) are examples of the fits using Eq.~(1) with the single singlet-triplet gap $\Delta_1$ = 5 K, where we assume the inner-AF-dodecamer excitations from the ground state with $S_{tot}=0$ to the excited state with $S_{tot}=1$; if we include only $\Delta_1$ in Eq.~(1), the softening in the experimental $C_{\Gamma}(T)$ for ZnFe$_2$O$_4$ cannot be reproduced. The gradient of the softening in $C_{\Gamma}(T)$ produced by Eq.~(1) becomes steeper at low temperatures and more gentle at high temperatures than the experimental data. Hence the experimental $C_{\Gamma}(T)$ for ZnFe$_2$O$_4$ suggests that the molecular-spin excitations consist of multiple gapped modes.

The level scheme of the AF-dodecamer excitations are not clarified so far, which should include inner- and outer-AF-dodecamer excitations. However, by assuming the coupling of the lattice to not only the lowest excitations $\Delta_1$ but also the higher excitations $\Delta_i$ ($i$ = 2, 3, 4, $\ldots$), Eq.~(1) reproduces well the experimental $C_{\Gamma}(T)$ for ZnFe$_2$O$_4$. Solid curves in Fig. 3(a)-(c) are examples of the fits using Eq.~(1) with the lowest singlet-triplet excitations $\Delta_1$, and the next higher singlet-triplet excitations $\Delta_2$. This assumption is similar to that applied to the AF-hexamer excitations in $A$Cr$_2$O$_4$.\cite{Watanabe} Given the values of the fitting parameters listed in the upper column of Table I, the fits to Eq.~(1) are in excellent agreement with the experimental data of ZnFe$_2$O$_4$ (Fig. 3(a)-(c)), reproducing the softening-with-minimum in $C_{\Gamma}(T)$. We note here that the fitted curves obtained by assuming $\Delta_2$ to be the singlet-quintet, -septet, and -nonet gaps, respectively, also give excellent agreement with the experimental data of ZnFe$_2$O$_4$ (the fitted curves are not shown in Fig. 3(a)-(c)). Thus, although the present study reveals the coupling of the lattice to the multiple AF-dodecamer excitations for ZnFe$_2$O$_4$, the nature of the excitations $\Delta_2$ cannot be identified so far.

In the fitting of Eq.~(1) to the experimental data of ZnFe$_2$O$_4$, we find that the magnitudes of $G_{1,\Gamma}$, $G_{2,\Gamma}$, and $|K_{\Gamma}|$ increase with increasing the degree of $\Delta_2$ degeneracy. This is exemplified by the values of the fitting parameters with the singlet-nonet $\Delta_2$ shown in parentheses in the upper column of Table I. On the other hand, we also find three qualitative features of $G_{1,\Gamma}$, $G_{2,\Gamma}$, and $K_{\Gamma}$, which are common regardless of the degree of $\Delta_2$ degeneracy and thus should be intrinsic features for ZnFe$_2$O$_4$. First, the $K_{\Gamma}$ values are negative for all the elastic modes, indicating that the inter-AF-dodecamer interaction is antiferrodistortive. Second, the coupling constant $G_{2,\Gamma}=|\partial \Delta_2/\partial \epsilon_{\Gamma}|$ is larger than $G_{1,\Gamma} =|\partial \Delta_1/\partial \epsilon_{\Gamma}|$, indicating that the higher excitations $\Delta_2$ couple to the dynamical lattice deformations more strongly than the lowest excitations $\Delta_1$. The larger value of $G_{2,\Gamma}$ than $G_{1,\Gamma}$ might suggest the coupling of the lattice to the higher multiple excitations $\Delta_i$ ($i$ = 2, 3, 4, $\ldots$). Third, both the coupling constants $G_{1,\Gamma}$ and $G_{2,\Gamma}$ exhibit the largest values for the trigonal shear modulus $C_{44}(T)$ of the three elastic moduli, and hence are compatible with the trigonal symmetry of the AF dodecamer.

In the lower column of Table I, the values of the fit parameters in Eq.~(1) for the experimental $C_{44}(T)$ of $A$Cr$_2$O$_4$ ($A$ = Mg and Zn) are also listed for comparison.\cite{Watanabe} For $A$Cr$_2$O$_4$, the value of $N$ in Eq.~(1) is assumed to be $N$ = 3.45$\times$10$^{27}$ m$^{-3}$ (one-sixth of the density of Cr$^{3+}$ ions in $A$Cr$_2$O$_4$), where the spin molecule is assumed to be the Cr$^{3+}$ AF hexamer (Fig.~1(e)).\cite{Lee2,Tomiyasu,Watanabe} As described before in conjunction with Fig.~2(a)-(c) and Fig.~2(f), the magnitude of the softening in $C_{\Gamma}(T)$ for ZnFe$_2$O$_4$ is smaller than that in $C_{44}(T)$ for MgCr$_2$O$_4$. In accordance with Eq.~(1), this difference in the magnitude between ZnFe$_2$O$_4$ and $A$Cr$_2$O$_4$ arises from the difference in the coupling strength between the dynamical lattice deformations and molecular-spin excitations. From Table I, the coupling constants $G_{1,\Gamma}$ and $G_{2,\Gamma}$ for $C_{44}(T)$ of ZnFe$_2$O$_4$ are smaller than those of $A$Cr$_2$O$_4$. Additionally, along with Figs.~3(a)-(c) and 2(f), $C_{\Gamma}(T)$ for ZnFe$_2$O$_4$ exhibits a minimum at $\sim$5~K, which is lower than the minimum point of $\sim$50~K in $C_{44}(T)$ for MgCr$_2$O$_4$. From Eq.~(1), this difference in the minimum point between ZnFe$_2$O$_4$ and $A$Cr$_2$O$_4$ arises from the difference in the magnitudes of the gaps $\Delta_1$ and $\Delta_2$ in the molecular-spin excitations.

For ZnFe$_2$O$_4$, the magnitudes of the inter-spin-molecule interactions $|K_{\Gamma}|$ listed in Table I are comparable to those of $\Delta_1$ and $\Delta_2$, which is in contrast to the weaker magnitudes of $|K_{\Gamma}|$ than $\Delta_1$ and $\Delta_2$ for $A$Cr$_2$O$_4$. The comparable magnitudes of $|K_{\Gamma}|$, $\Delta_1$, and $\Delta_2$ for ZnFe$_2$O$_4$ imply that the inter-spin-molecule interactions are not completely negligible, which is compatible with the presence of a very weak dispersive feature of the molecular-spin excitations in the INS spectra of ZnFe$_2$O$_4$, suggesting the presence of the inter-spin-molecule correlations.\cite{Kamazawa} Taking into account the smaller values of $\Delta_1$ and $\Delta_2$ for ZnFe$_2$O$_4$ than for $A$Cr$_2$O$_4$, the very weak dispersion observed in the INS spectra of ZnFe$_2$O$_4$ might be a result of frustration occurring in the effectively weaker exchange interactions.

We finally note that the recent time-of-flight INS experiments in the PM phase of MgCr$_2$O$_4$ revealed the presence of multiple modes associated with finite-energy molecular-spin excitations, where it is suggested that not only the ground state but also the excited states are highly frustrated in this compound.\cite{Tomiyasu2} Hence, although the assumption of two gaps $\Delta_1$ and $\Delta_2$ gives from Eq.~(1) a successful agreement with experimental data for $C_{\Gamma}(T)$ of both ZnFe$_2$O$_4$ and $A$Cr$_2$O$_4$, it is expected that the level schemes of the spin molecules in these compounds consist of not only the excited levels of $\Delta_1$ and $\Delta_2$ but also higher excited levels. Similar to MgCr$_2$O$_4$, the future time-of-flight INS experiments in ZnFe$_2$O$_4$ are expected to reveal complex and exotic molecular-spin excitations.

\section{Summary}

In summary, ultrasound velocity measurements of ZnFe$_2$O$_4$ revealed the elastic anomalies, which strongly suggest the evolution of the molecular-spin excitations at low temperatures. Additionally, the present study also revealed that the elastic anomalies driven by the magnetostructural phase transition and its precursor are absent in ZnFe$_2$O$_4$, suggesting that the spin-JT mechanism cannot play a role in releasing frustration within this compound. The present study infers that, for ZnFe$_2$O$_4$, the dynamical molecular-spin state evolves at low temperatures without undergoing precursor spin-lattice fluctuations and spin-lattice ordering. Further experimental and theoretical studies are indispensable if the dynamical molecular-spin state in ZnFe$_2$O$_4$, which is expected to govern the magnetic properties, is to be understood.

\section{Acknowledgments}

This work was partly supported by a Grant-in-Aid for Scientific Research (C) (Grant No. 25400348) from MEXT of Japan, and by Nihon University College of Science and Technology Grants-in-Aid for Project Research.

\end{document}